# Mapping the Chinese Science Citation Database

## Loet Leydesdorff

University of Amsterdam, Science & Technology Dynamics, Amsterdam School of Communications Research (ASCoR), Kloveniersburgwal 48, 1012 CX  Amsterdam, The Netherlands. Email: loet@leydesdorff.net

## Jin Bihui

Library of the Chinese Academy of Sciences, 33 Beisihuan Xilu, Zhongguancun, Beijing 100080, P. R. China. Email: jinbh@mail.las.ac.cn .

**Methods developed for mapping the journal structures contained in aggregated journal-journal citations in the *Science Citation Index* are applied to the *Chinese Science Citation Database* of the Chinese Academy of Sciences. This database covers 991 journals, of which only 37 had originally English titles. Using factor-analytical and graph-analytical techniques we show that this data is dually structured. The main structure is the intellectual organization of the journals in journal groups (as in the international *SCI*), but the university-based journals provide an institutional layer that orients this structure towards practical ends (e.g., agriculture). The *Chinese Science Citation Database* exhibits the characteristics of "Mode 2" in the production of scientific knowledge more than its western counterparts. The contexts of application lead to correlation (interfactorial complexity) among the components.**

## Introduction

Aggregated journal-journal citation relations have been organized in the *Journal Citation Reports* of the *Science Citation Index* on a yearly basis since 1975. As early as 1965, on the basis of an experimental version of this database, Derek de Solla Price noted the pronounced structure of this matrix and he suggested that it would be possible to decompose it for the mapping of scientific specialties and disciplines (Price, 1965; cf. Simon, 1969). The time series allows us additionally to study the dynamics of the sciences (Leydesdorff, 2002).

At the level of journal clusters, the dynamics can be considered as a baseline that is relatively independent of intentional (social or political) agency (Studer & Chubin, 1982; Zsindely *et al.*, 1982; Leydesdorff, 1987). In national contexts, however, one can wonder whether a more direct couplings between institutional interests and journal structures might exist. Perhaps national elites provide an intermediating, but invisible college (Crane, 1972; Mulkay, 1976). For example, the French CNRS subsidizes approximately 225 journals which are only partially covered by the U.S.-based *Science Citation Index* (De Looze *et al.*, 1996; Legentil, personal communication). However, Sivertsen (2003) found no bias in the coverage of Scandinavian publications by the *Science Citation Index* (Garfield, 1979, 1990).

The Institute of Scientific Information (ISI) admits that the selection system for inclusion in the *Science Citation Index* may be biased against journals written in non-latin alphabets. Special care has been taken in the past to include Russian and Japanese journals into the database (Garfield, 1979, 1998). Chinese scientific journals, however, have not been systematically evaluated for their inclusion in the ISI database. The expanded version of the *SCI* at the Internet included only 31 Chinese scientific journals in 2001. Jin *et al.* (2002) estimated this as 0.73% of the scientific journal titles available in China.

The Library of the Chinese Academy of Sciences has developed the *Chinese Science Citation Database (CSCD)* since 1989 (Jin & Wang, 1999; cf. Liang & Wu, 2001). In this project, the data for the year 2001 was aggregated in a format similar to the *Journal Citation Reports* of the *SCI*. We apply graph-analytical and factor-analytical techniques that were developed in previous projects for analyzing the international *Science Citation Index* to the Chinese dataset (Leydesdorff & Cozzens, 1993; Leydesdorff, 2004a).

## Methods and Materials

### *Data*

We have examined aggregated journal-journal citation data for 991 Chinese journals. Only 37 of these journals have titles originally in English. The other 954 journal titles were translated into English for the purpose of this project. Among these journals 55,774 citation relations are maintained, that is, 5.7% of the 982,081 (= $991^2$) possible relations. The corresponding figure was 2.6% for the *Science Citation Index* and 2.8% for the *Social Science Citation Index* in the year 2001. However, these indices exclude most of the single citation relations (using a threshold). Since the single citation relations amount in the *CSCD* to 28,454, the Chinese figure corrected for this comparison would be 2.8%. We pursue the analysis below, including these single occurrences in the Chinese citation matrix.

### *Methods*

The matrix of 991 x 991 cells was first constructed and then saved as an SPSS systems file. This file can, for example, be factor analyzed. We will use both the overall factor analysis and the routines which we previously developed for the local analysis of journal-journal structures (Cozzens & Leydesdorff, 1993; Leydesdorff, 1986; Leydesdorff & Cozzens, 1993).

In addition to factor analysis we use graph-analytical techniques from social network analysis (Otte & Rousseau, 2002). In particular, bi-connected component analysis has recently been further developed. This technique was originally developed in order to identify robust clusters in large data sets (Knaster & Kuratowski, 1921) and was more recently incorporated in software tools for social network analysis (Moody & White, 2003).

A bi-connected component provides us with a robust definition of a cluster because the network component remains connected after removing any one of the vertices (Mrvar & Bagatelj, s.d.). A network component is called *bi-connected* if for every three vertices *a*, *v*, and *w* there exists a chain between *v* and *w* which does not include the vertex *a*. The bi-connectedness stabilizes the clusters, for example, against changes and variations in the initial selection when producing the database. Thus, the incidental inclusion or exclusion of journals does not directly affect the categorization of the network data.

In a previous study Leydesdorff (2004a) analyzed the 5,748 journals of the *Science Citation Index 2001* in terms of this algorithm. 3,991 (that is, 73.8%) of the journals were related in a single network at the relatively high level of correlation of $r = 0.8$ among the citing patterns. In the *Social Science Citation Index 2001* the corresponding figure was only 781 or 48.6% of the total of 1608 journals included. In this latter case, therefore, we increased the value of the Pearson correlation coefficient stepwise from 0.2 with increments of 0.1 (Leydesdorff, 2004b). An analogous procedure is used in this case because exploration of the data has taught us that some bi-connected components can already be extracted from this dataset at the relatively low level of $r = 0.2$. (Only 62 journals can be organized in 17 bi-connected components at the high threshold level of $r \geq 0.8$.)

## Results

### *Factor analysis at the system level*

The matrix of 991 citing journals can be decomposed using factor analysis. Nine factors have an eigenvalue larger than unity and thus explain more than an average variable. These first nine factors explain 15.47% of the common variance.[1] However, extraction of these nine factors provided us with a mixture of positive and negative factor loadings on the last (ninth) factor. By allowing for ten factors, the factor analytical solution became unambiguous. Table 1 shows the three journals with highest factor loading for the first three factors as an example.

Table 1. Three journals with highest factor loadings for each of the first three factors in a ten factor solution.[2]

Rotated Component Matrix

| | 1 | 2 | 3 | 4 |
|---|---|---|---|---|
| *Sci Agr Sin* | .793 | | | |
| *Acta Agronomica Sin* | .751 | | | |
| *Chin J Oil Crop Sci* | .739 | | | |
| *Chin J Geol* | | .871 | | |
| *Bull Miner, Petrol Geochem* | | .862 | | |
| *Earth Sci* | | .859 | | |
| *J Harbin Med Univ* | | | .747 | |
| *J Chongqing Med Univ* | | | .738 | |
| *Jiangsu Med J* | | | .738 | |

The ten factors can be designated in terms of disciplinary affiliations. They also reflect the priorities of the Chinese economy, which is still mainly resource-based. The journals loading on the third factor exhibit a local component by being tied to specific medical schools, while the more specialized journals follow with lower factor loadings. For example, the *Chinese Journal of Internal Medicine* obtains only the 16th position on this list with a factor loading of 0.584. In the *Science Citation Index* the specialist journals prevail in the biomedical domain.

[1] In the case of the *Social Science Citation Index* 12 eigenvalues are larger than unity, and these twelve factors explain 26.29% of the variance. (Nine factors explain 22.37%.) This analysis could not be pursued with the full set of 5,748 jouranls of the *Science Citation Index* because of systems limitations (in the subroutine for factor analysis), but the expectation is that the eigenstructure of this database is far more pronounced than the *Social Science Citation Index* (Leydesdorff, 2003).

[2] Extraction Method: Principal Component Analysis. Rotation Method: Varimax with Kaiser Normalization; rotation converged in 9 iterations.

Furthermore, unlike chemistry (factor 5), mathematics (factor 4), and geology (factor 2), the physics journals do not appear as a separate grouping. Titles of physics journals are subsumed under the various applications of physics such as material sciences (factor 9), advanced optics (factor 10), and computer engineering (factor 6). We shall see below that this focus on the context of application is typical for the organization of these journals.

When the (symmetrical) matrix is transposed in order to study the Q-structure in the cited dimension, we find essentially the same factors, but the order is different. A ‘pharmacology’ factor replaces one of the previous factors. Table 2 summarizes these results.

In summary, we found a disciplinary structure in the citation structures of the *Chinese Science Citation Database*, both in the cited and the citing dimensions. Institutional structures related to specific universities are visible in citation patterns among journals in the medical sciences. This suggests that in the publication and citation patterns, the clinical side prevails over the research side. Similarly, in physics, we noted the dominance of the applied side. Otherwise, the intellectual codification structures the differentiation among the journals into sets. In the next section, we investigate the 10+ disciplines distinguished above in more detail by using a graph-analytical approach.

Table 2. Factor structure of the ten first factors in the citing and the cited dimension of the matrix.

| | *Citing* | *Cited* |
|---|---|---|
| *Factor 1* | Agriculture | Medicine |
| *Factor 2* | Geology | Agriculture |
| *Factor 3* | Medicine | Geology |
| *Factor 4* | Chemistry | Comput. Eng. |
| *Factor 5* | Mathematics | Environ. Sci. |
| *Factor 6* | Comput. Eng. | Chemistry |
| *Factor 7* | Environ. Sci. | Mat Sci Tech |
| *Factor 8* | Oceanology | Mathematics |
| *Factor 9* | Mat Sci Tech | Pharmacy |
| *Factor 10* | Optoelect | Oceanology |

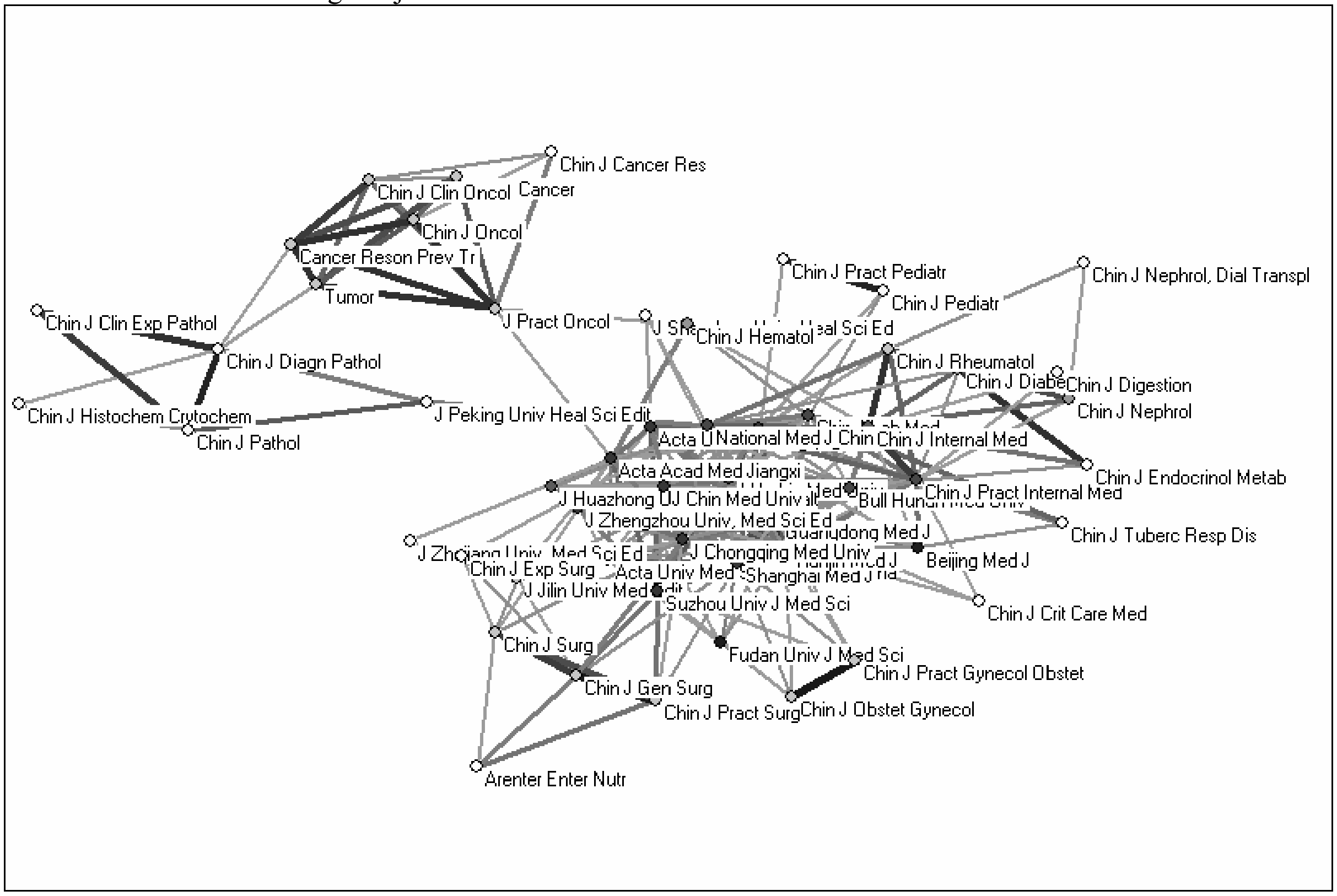

Figure 1. Bi-connected component of 54 journals in medicine that relate at the level $r \geq 0.4$

### *Bi-connected components*

The matrix can be decomposed into 70 bi-connected components. These components vary in terms of the internal strength of the citation relations. For example, the largest group is composed of a set of 54 journals in medicine which relate at the level of $r \geq 0.4$. This bi-component is visualized in Figure 1. The other factors mentioned in Table 2 equally provide bi-connected components with a number of journals included. However, one can also find similar components which do not belong to these central groupings. (The complete decomposition can be retrieved at http://www.leydesdorff.net/china01. )

Figure 1 confirms the suggestion about the degree of specialization in medical journals contained in this set. Separate groupings of oncology and pathology journals are visible, but the core group is organized in terms of general and clinical medicine. Note that this component cannot be found when using the international *Science Citation Index* for this analysis because journals belonging to the medical specialties prevail in the citation patterns in this database (Leydesdorff, 2004a; see for the complete decomposition at http://www.leydesdorff.net/jcr01 ).

Table 3. Factor structure among citing patterns of 27 journals in the citation environment of *Scientia Agricultura Sinica* (Threshold at 1% of total cited or total citing of the seed journal.)

| | Component | | | | | | | |
|---|---|---|---|---|---|---|---|---|
| | 1 | 2 | 3 | 4 | 5 | 6 | 7 | 8 |
| Chin J Rice Sci | .878 | | | .124 | | .105 | .195 | |
| Acta Agronomica Sin | .871 | .234 | | .156 | .217 | -.115 | .252 | |
| J Yangzhou Univ-Agr Life Sci | .841 | | | .132 | .198 | -.102 | .210 | |
| Southwest Chin J Agr Sci | .805 | | .236 | | | .141 | -.117 | |
| Sci Agr Sin | .766 | .208 | .248 | .364 | .164 | .226 | .130 | .138 |
| J Sichuan Agr Univ | .740 | | | | | .170 | -.191 | |
| J Triticeae Crop | .691 | .218 | -.114 | .318 | .346 | -.270 | .304 | |
| Chin J Oil Crop Sci | .675 | .312 | | .389 | .252 | | .141 | .412 |
| Acta Genet Sin | .600 | .168 | -.239 | | .102 | | | -.212 |
| J Hunan Agr Univ | .489 | | .281 | | -.221 | | | |
| Acta Agr Boreali Sin | .446 | .235 | | .184 | .380 | -.305 | .365 | |
| J Fujian Agr Forest Univ | .370 | .136 | .200 | | -.288 | | | -.105 |
| Chin Bull Bot | .151 | .888 | | | | | | |
| Acta Bot Boreali-Occidentalia | .155 | .865 | .122 | | .155 | | | |
| Plant Physiol Commun | | .833 | .367 | .114 | | | | |
| Acta Hortic Sin | | .177 | .939 | | | | | |
| J Fruit Sci | | .288 | .931 | | | | | |
| Acta Phytopathol Sin | | | | .889 | | | | |
| J Shenyang Agr Univ | .249 | .167 | .226 | .840 | | .123 | | |
| J Shandong Agr Univ | .402 | .119 | .510 | .511 | .298 | -.143 | .178 | |
| Acta Agr Boreali-Occidentalis | | | | | .768 | | | |
| Plant Nutr Fertilizer Sci | .358 | | .118 | | .446 | .116 | | |
| Hubei Agr Sci | .419 | | .209 | .168 | | .679 | .117 | |
| Cotton Sci | | | -.145 | .418 | .141 | .566 | .251 | -.212 |
| J Jilin Agr Univ | .186 | -.132 | | .230 | | -.380 | | |
| J Anhui Agr Sci | .120 | | | | -.108 | .192 | .839 | |
| Boybean Sci | | | | | | | | .920 |

Extraction Method: Principal Component Analysis. Rotation Method: Varimax with Kaiser Normalization.
a Rotation converged in 11 iterations.

Table 3 shows that the (first) grouping of agricultural journals is internally composed of different subgroupings which do not relate to each other in terms of the bi-connected graph analysis. Twelve agricultural journals which stand at the top of the list of factor loadings in the first factor at the system level, form one bi-component at the threshold level of r ≥ 0.6. However, Figure 2 provides a visualization of a larger group of 28 journals in botany which are internally related at this same level of correlation.

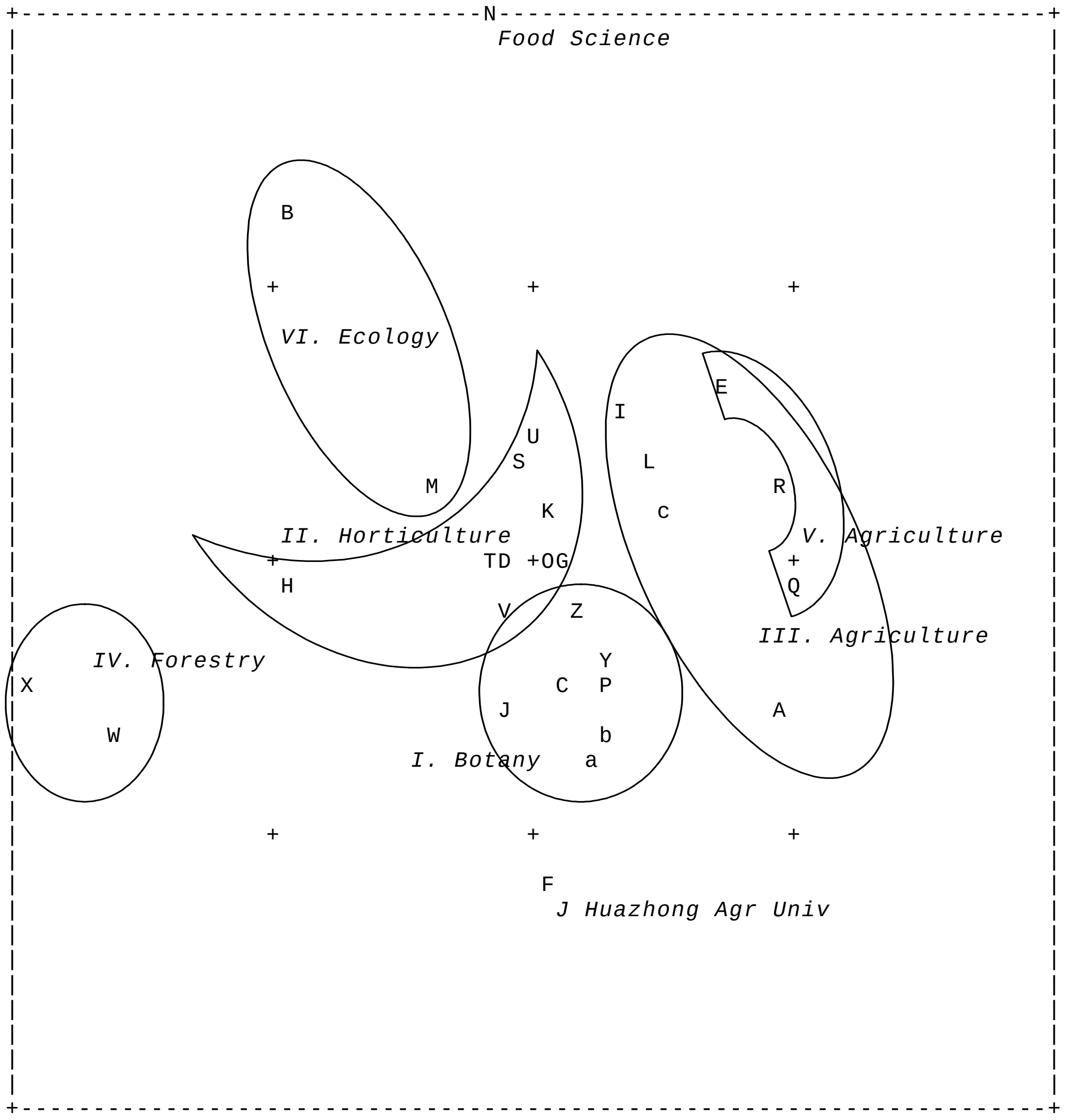

Figure 2. Multidimensional Scaling of 29 journals in the citation environment of *Acta Horticulturae Sinica*

A. J Fujian Agr Forest Univ
B. Guangdong Agr Sci
C. Guihaia
D. J Fruit Sci
E. J Hunan Agr Univ
F. J Huazhong Agr Univ
G. J Laiyang Agr Coll
H. J Lanzhou Univ
I. Liaoning Agr Sci
J. Chin J Trop Crops
K. J Shandong Agr Univ
L. J Shenyang Agr Univ
M. Ecol Sci
N. Food Sci
O. J Northwest Sci Tech Univ Agr
P. Acta Bot Boreali-Occidentalia
Q. J Southwest Agr Univ
R. Southwest Chin J Agr Sci
S. Xinjiang Agr Sci
T. Acta Hortic Sin
U. J Yunnan Agr Univ
V. J Zhejiang Univ, Agr Life Sci
W. J Zhejiang Forest Coll
X. Zhejiang Forest S&T
Y. J Plant Physiol Mol Biol
Z. Plant Physiol Commun
a. Acta Bot Sin
b. Chin Bull Bot
c. Sci Agr Sin

Similar differentiations into fine-grained disciplinary structures of specialties can be distinguished in the other major dimensions of the matrix. The bi-connected components are comparable with those in the *Science Citation Index* and the *Social Science Citation Index*, but the Pearson correlations are on average lower within the clusters. In the next section, we use factor analysis at this lower level of aggregation for the analysis of these correlation matrices.

### *Factor analysis at the specialty level*

Table 3 shows the (default) factor structure using the journal *Scientia Agricultura Sinica* as the seed journal for the generation of a local citation environment. This journal was the one with the highest factor loading on the first factor at the system level. The factor was designated above as representing "agricultural sciences."

The component matrix shows that the 27 journals which are citing or cited by this seed journal at more than one percent of its total citation rate in 2001 are organized in eight dimensions (with eigenvalues $\geq$ 1). The first grouping represents "agricultural sciences," the second one "botany," and a third one "horticulture." Smaller groupings are also distinguished. These are sometimes strongly related to regional centers at specific universities.

Most remarkable about this matrix is the considerable filling of the "off-diagonal" cells. Although the highest factor loading on the main dimensions are at levels of $r \geq 0.8$, the interfactorial complexity cannot be neglected. This is only seldom the case in this type of analysis when using data from the *(Social) Science Citation Index*. It indicates that the various dimensions of the intellectual organization (which prevails) are also organized in another dimension, notably the institutional one of regional and national universities.

The second factor (with highest factor loading for the *Chinese Bulletin of Botany*) indicates a group of botany journals, but the intellectual organization of botany is not well represented from this (agricultural) perspective on the journal structure. The journal *Acta Botanica Sinica*, for example, which is also included in the international *Science Citation Index,* does not play a significant role in this citation environment.

If we approach the journal structure from this latter angle using *Acta Botanica Sinica* as a seed journal, however, we lose the major agricultural journals. Thus, the divide between the intellectual organization and the applied-science side can be considerable. We need to use a horticultural journal (*Acta Horticulturae Sinica*) as a point of entry for bringing the two fields into a single perspective (Figures 2 and 3).

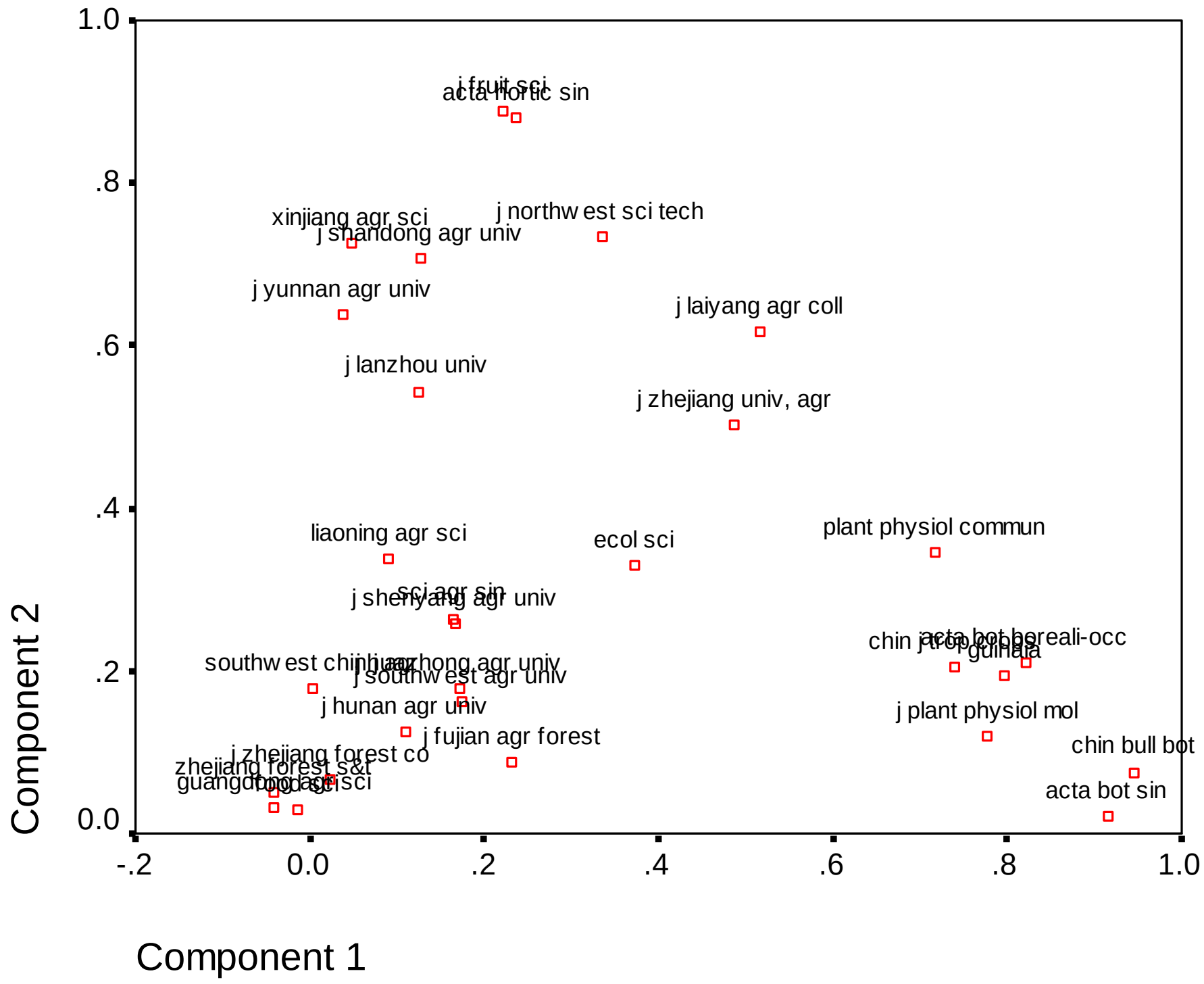

Figure 3. Factor plot of two main components in the citation environment of *Acta Horticulturae Sinica*

.

Figure 3 provides a factor plot of the two main factors ("agriculture" and "botany") when analyzed from the perspective of "horticulture". The figure shows that the journals are not only positioned along the main axes of the system, but that the interfactorial complexity provides some journals with mediating roles (Van den Besselaar & Heimeriks, 2001). These journals are often associated with specific universities.

## Conclusions

While the *Science Citation Index* and the *Social Science Citation Index* exhibit the prevailing differentiation among journal groupings in terms of specialties and disciplines, the *Chinese Science Citation Database* is additionally integrated by a maze of university-based journals. The latter are sometimes also specialized, for example, in the medical field. The main organization of these university-based journals, however, is not determined by their intellectual or institutional affiliations, but by the context of application of these journals. In general, these journals have lower impact factors than the nation-wide journals.

We elaborated this tension between differentiation and integration in the case of the relation between journals in botany and agriculture above, but the pattern was also indicated in medicine and in physics. The layer of university-based journals generates inter-factorial complexity among the otherwise intellectually organized dimensions of the citations in the journal set. In a next study, we intend to test this hypothesis by using oblique rotation in the factor analysis when comparing journals in the overlap between the *Chinese Science Citation Database* and the *Science Citation Index.*

In summary, one could say that the *Chinese Science Citation Database* exhibits the characteristics of "Mode 2" in the production of scientific knowledge more than its western counterparts. Gibbons *et al.* (1994) suggested that the contexts of application are increasingly structuring the organization of international science in a transdisciplinary mode. These authors labeled this as "Mode 2." While in "Mode 1" science intellectual organization would mainly codify the communication, e.g., through peer review, in "Mode 2" science selections would increasingly be based on heuristics with reference to applications.

In a more recent study, Nowotny *et al.* (2001) argued that the contextualization does not only affect the institutional parameters (e.g., missions), but that the intellectual organization of the sciences itself is increasingly reorganized. Although there remains little evidence of an emerging Mode 2-type of science when using scientometric indicators (Leydesdorff, 2001; Shinn, 2002), the organization of the Chinese journals exhibit the applicational contexts as relevant albeit as a secondary structure in addition to the intellectual organization. This layer of institutional integration provides a focus on the priorities of the economy and the state that can be considered absent in the international database, while the latter is mainly differentiated in terms of specialties and disciplinary structures. However, with the rapid internationalization of Chinese science in recent years (Jin & Leydesdorff, in preparation), one may expect this secondary structure to become less important in the near future.

## AKNOWLEDGEMENTS

The authors wish to thank Zhang Wang for the organization of the data.